\documentclass[final,3p,times,twocolumn]{elsarticle}
\usepackage{amssymb}
\journal{Physica C}

\begin{document}

\begin{frontmatter}



%
%

\title{Clustering Superconductors Using Unsupervised Machine Learning}

%
%
%
\author{B. Roter$^{*}$}
\address{Department of Physics, The University of Akron,
Akron, OH 44325, USA}
\address{$^{*}$Present address: Program of Applied Physics, 
Northwestern University, Evanston, IL 60208, USA}
\author{N. Ninkovic}
\address{Department of Physics, The University of Akron,
Akron, OH 44325, USA}
\author{S.V. Dordevic}
\address{Department of Physics, The University of Akron,
Akron, OH 44325, USA}
\ead{dsasa@uakron.edu}

\date{\today}

%
%
\begin{abstract} 
In this work we used unsupervised machine learning methods in order
to find possible clustering structures in superconducting 
materials data sets. 
We used the SuperCon database, as well as our own data sets complied 
from literature, in order to explore how machine learning algorithms 
groups superconductors. Both conventional clustering methods 
like k-means, hierarchical or Gaussian mixtures, as well as clustering 
methods based on artificial neural networks like self-organizing 
maps, were used. For dimensionality reduction and visualization 
t-SNE was found to be the best choice. 
Our results indicate that machine learning techniques can achieve, 
and in some cases exceed, human level performance. Calculations
suggest that the clustering of superconducting materials works best
when machine learning techniques are used in concert with 
human knowledge of superconductors. We also show that in order 
to resolve fine subcluster structure in the data, clustering of 
superconducting materials should be done in stages.
\end{abstract}

%
%
%






\end{frontmatter}

\section{Introduction}

In recent years a number of artificial intelligence (AI) studies of superconductors 
have been performed \cite{stanev18,hamidieh18,konno18,matsumoto19,zeng19,roter20,liu20},
using either machine learning (ML) or deep learning techniques. 
In some of these studies, the superconducting critical 
temperature of known superconductors was predicted, i.e. calculated. 
Most of these studies achieved impressive values of 
statistical parameters, such as R$^2$ or RMSE. However, none
of them have (yet) led to the discovery of new superconducting 
materials, which should be the ultimate goal of AI studies.

It was recently pointed out \cite{meredig18} that AI models achieved 
high values of statistical parameters because they used traditional 
k-fold cross-validation procedures, in particular leave-one-out cross 
validation which is its special case. These cross validation
techniques have been shown to result in unrealistically 
high values of statistical parameters when applied to highly 
clustered data sets \cite{meredig18}. Moreover, they have been known 
to have problems when extrapolating to new classes of materials 
\cite{stanev18}. As an alternative, an approach called leave-one-cluster-out 
cross validation was proposed \cite{meredig18}, which was supposed
to alleviate the problem. The idea is that instead of searching
for new superconducting materials, AI is used to search for 
new {\it classes} of superconducting materials. 
However, the problem is that there 
is currently no universally accepted classification of superconductors. 
There have been numerous attempts to classify superconductors 
based on their crystal structure, physical properties, pairing 
mechanics, etc. \cite{poole-book,hirsch15} but none of them are
universally accepted.

Over the past century, superconductivity has been found in a variety
of different systems, such as metals, oxides, alloys, organic 
materials, heavy fermions, etc. These systems can have vastly different 
physical, chemical and all other properties, which makes it difficult 
for humans to classify.  In a recent study Hirsch, Maple and
Marcillio \cite{hirsch15} made the most comprehensive
attempt yet to separate superconductors into different classes, 
based on the physical mechanism of superconductivity. 
They identified 32 classes, separated into three big groups:
conventional (with 12 classes), potentially unconventional (with 9 
classes) and unconventional (with 11 classes). However, some 
superconductors in this study can be found in several different
classes.  

In this work we used a ML technique called clustering in order to 
separate superconductors into classes (or clusters). Our hope is
that ML might be able to separate superconductors into 
physically meaningful groups, which would in turn lead to 
the discovery of entirely new groups of superconductors. Additionally,
ML techniques
have been known to both provide additional insights and reveal hidden 
patterns in other groups of materials not previously
known to humans \cite{hill16}.
 
Our results reveal that data-driven ML techniques can indeed separate superconductors
into groups, achieving and in some cases exceeding human level performance. 
However, there are also limitations, which will need to be addressed in 
the future. In particular, constructing a more reliable database, which
will undoubtedly result in better ML calculations. 
Clustering results obtained on the test data sets
we compiled from literature, confirm this claim. 
Our calculations indicate that, at least for superconducting materials data,
clustering in stages is the best approach.


\section{Clustering}

Clustering is one of the most common tasks of unsupervised 
machine learning \cite{wunsch-book,kogan-book}. 
The main goal of unsupervised learning algorithms is to find patterns 
and learn meaningful relationships in data in order to describe its 
underlying structure. However, in unsupervised machine learning one does 
not know the correct output; thus the results of 
calculations cannot be assessed in terms of accuracy or correctness. 
Instead, clustering is an exploratory technique. One must not assume 
there is a single "true" clustering. One should rather be 
interested in exploring different clusterings of the same 
data set to learn more about it. Using commercially available clustering 
programs as a "black box" should not be done, as it may lead
to erroneous results and conclusions. 

Before we attempt to cluster superconductors, there are several 
issues that need to be addressed, such as the choice of a clustering 
technique, the number of clusters, the choice of the distance function, 
dimensionality reduction and visualizations.
All these issues are discussed below in separate subsections
before clustering results are presented.

\subsection{Conventional Clustering Methods}

We used several conventional clustering methods 
\cite{wunsch-book,kogan-book}: k-means, k-medoids, hierarchical, 
Gaussian mixtures, DBSCAN, etc. For most of them
the number of clusters is an input parameter, which is selected
based on the data set. DBSCAN is one of the few that does not 
require the number of clusters to be specified. 
Another important factor that affects clustering is the distance 
function used to calculate similarly/dissimilarity between data points.
We have tried several standard distance functions such as 
Euclidean, cityblock, cosine, Chebychev, etc. In our calculations
with superconductors, the cosine distance function has usually produced 
the best results. 

A procedure for selecting the best clustering method was recently
proposed in Ref.~\cite{parker19}. It is referred to as Iterative Label 
Spreading, and it allows one to visually assess the quality of 
clusters, before any clustering is done. It was shown to be 
particularly suitable for noisy, high dimensional data
typically encountered in materials science.

\subsection{Neural Network--Based Clustering Methods}

In recent years artificial neural networks, especially deep 
neural networks, have become widely used in a number of 
disciplines such as image and speech processing, chemistry, biology,
material science, etc \cite{lecun15}. Although neural 
networks are used primarily for supervised learning tasks \cite{lecun15} 
such as regression and classification, they can also be used 
for unsupervised learning, i.e. clustering.

In this work we used a neural network based clustering method
known as the Self-Organizing Map (SOM) algorithm. 
SOMs perform nonlinear transformations of a multidimensional data set
into a low-dimensional set that retains the intrinsic topological
properties of the original data set. The network is trained to 
transform a so-called "input space", whose dimension equals the 
number of predictors, into a low-dimensional output called  "map space".
The most important parameter of these calculations
is the number of nodes, which is equivalent to the number of 
clusters. In addition, one also needs to carefully choose 
the topology of the input layer. Two most commonly used 
topologies are rectangular and hexagonal. In our calculations with
superconducting materials, SOMs have consistently performed 
better compared to conventional clustering algorithms.

\subsection{The Choice of Predictors}

One of the most important issues in machine learning calculations
is the choice of predictors. 
Similar to what we did in our previous work \cite{roter20},
we used chemical composition as the {\it only} predictor for 
clustering calculations. We previously showed that the use 
of a large number of predictors 
does not necessarily lead to any improvements in ML models. 
We suggested that the predictors must be carefully selected
based on the machine learning task. In calculations with 
superconductors, one should use those predictors that are known 
(or at least believed) to be closely related to superconductivity,
such as crystal symmetry, electronic band structure, number
of valence electrons, etc. 
The problem is that these predictors are
not systematically reported in the existing databases. They
should be included in future databases, as they might result
in more reliable machine learning calculations. 

\subsection{The Number of Clusters}

As mentioned above, clustering is an exploratory technique, 
and the "correct" number of clusters does not exist. Most
clustering techniques treat the number
of clusters as a parameter of calculations, which can be 
adjusted based on the data set. There are several 
algorithms one can use in order to estimate the optimal number 
of clusters. The most commonly used are:  
Calinski-Harabasz criterion, Davies-Bouldin index,
silhouette and dendrogram plots \cite{wunsch-book}. 
A dendrogram plot in particular allows one to explore 
not only cluster, but also sub-cluster structure 
in the data. Iterative Label Spreading can also be used 
to estimate the number of clusters \cite{parker19}. However, 
in our experience, the best way to estimate the optimal
number of clusters is by combining machine learning calculations
with human knowledge of superconducting materials. This approach
can produce the most physically meaningful clustering. 

Another important issue that one must be aware of is 
outliers in the data set used. 
In most AI studies of superconductors the SuperCon database \cite{supercon}
was used, which contains a number of wrong entries, either in 
terms of chemical composition or in terms of superconducting
critical temperature T$_c$. We previously estimated that
they might account for up to 20$\%$ of the entire database,
in particular for the cuprate entries \cite{roter20}.  
These wrong entries can have strong effect on AI calculations,
as they can create false clusters.
Future calculations would benefit from an improved database, as we
show here by using our own data sets compiled from literature.

\subsection{Dimensionality Reduction and Data Visualization}

Similar to what we did in our previous publication \cite{roter20}, 
we used element-vectors to mathematically represent superconductors
based on their chemical composition. In this approach, each superconductor 
is represented with a 1$\times$96 vector. As is sometimes done in machine 
learning, this dimension can be reduced to allow more efficient calculations. 
A variety of different techniques have been developed over the years:
PCA (or SVD), random projections, MDS, etc. 
\cite{matasov20}. However, our calculation with superconductors 
indicate that the best results are achieved when using a nonlinear 
dimensionality reduction technique called t-distributed 
Stochastic Neighbor Embedding (t-SNE) \cite{maaten08}. 

t-SNE has been used extensively for machine learning calculations in
materials science, and our calculations also indicate that it is the 
best choice for superconductors. It is a non-linear transformation 
that reduces the number of dimensions down to two or three, none of
which have any physical meaning. This not only allows
more efficient calculations, but also visualization
of high dimensional data in a low dimensional space. 
Another useful feature of t-SNE is that it allows easy manual 
backward validation, which we used extensively to investigate 
interesting data points, as well as outliers. 

\begin{figure}[tbp]
\vspace*{0.0cm}%
\centerline{\includegraphics[width=4in]{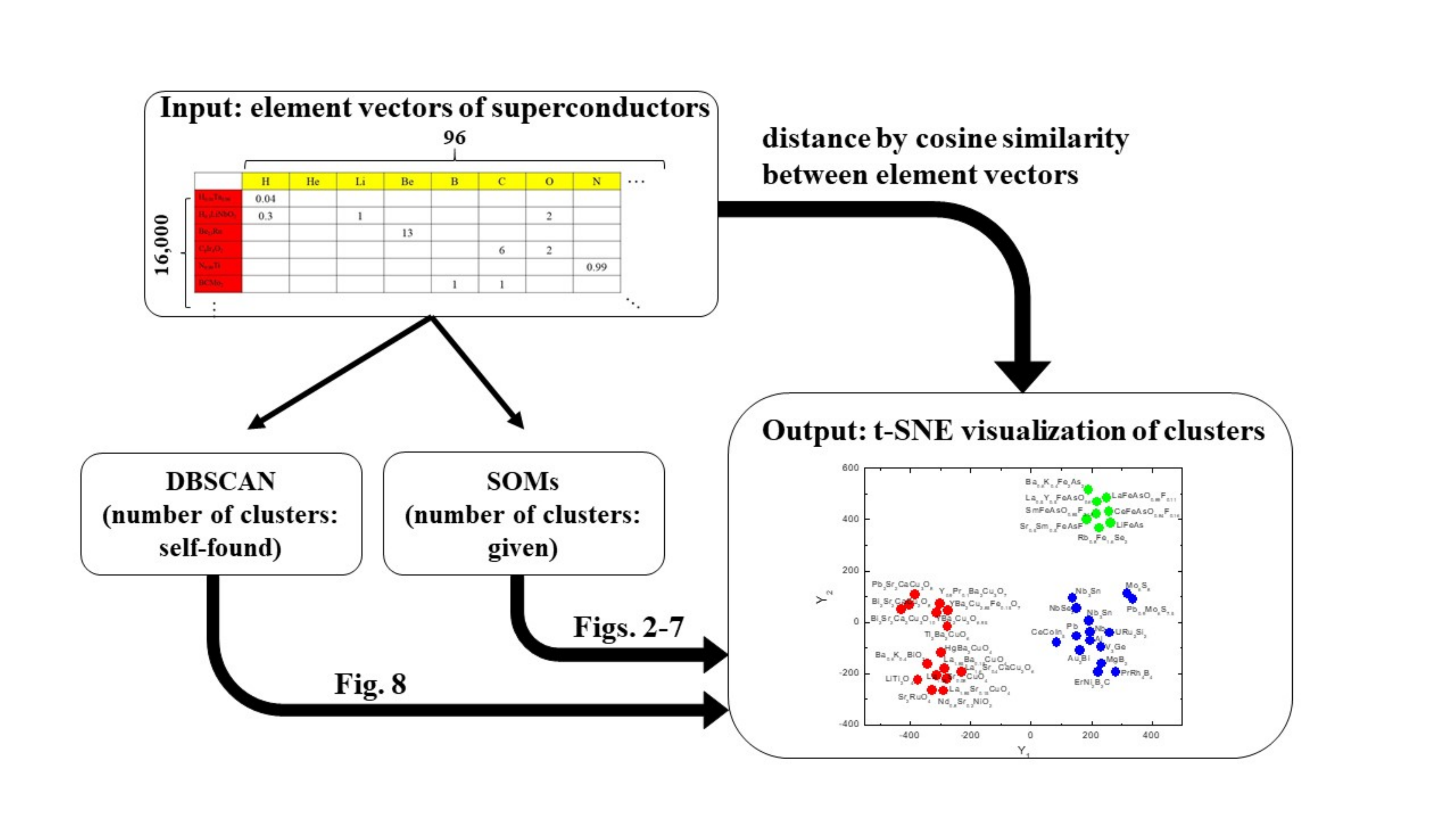}}%
\vspace*{-0.25cm}%
\caption{(Color online). The work flow chart for the calculations 
done in this project.}
\vspace*{0.0cm}%
\label{fig:workflow}
\end{figure}

One must be careful when using t-SNE, however. Due to its stochastic nature, 
its results can be difficult to interpret \cite{wattnberg16}.
The most important parameter for t-SNE calculations is the so-called 
perplexity. Perplexity value must be carefully chosen to avoid 
misinterpretations \cite{wattnberg16}. 
In our calculations we performed a number of runs with different values of 
perplexity in order to eliminate spurious features. Depending on the size
of analyzed data set, the values of perplexity we used were between 5 and 500. 
Fig.\ref{fig:workflow} shows schematically the work flow chart for the 
whole project.


\section{Clustering Test Data Sets}

\begin{figure}[tbp]
\vspace*{0.0cm}%
\centerline{\includegraphics[width=4in]{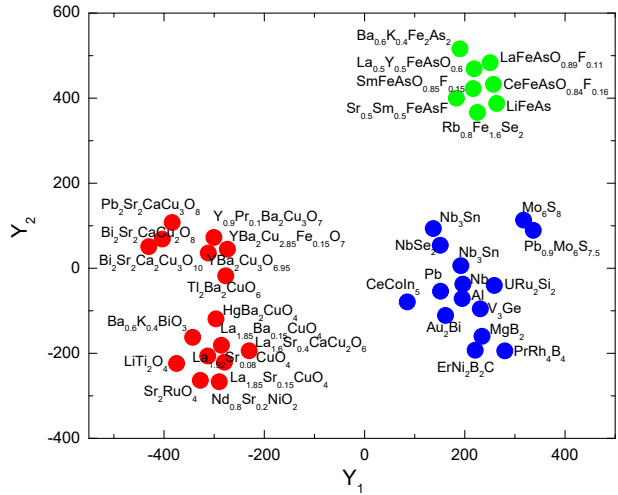}}%
\vspace*{-0.25cm}%
\caption{(Color online). t-SNE plot of the test set consisting 
of 37 superconductors complied from literature. The two axes
do not have any physical meaning. Three clusters can be identified and they 
correspond to cuprates (red circles), pnictides (green circles) and 
other (blue circles) superconductors.}
\vspace*{0.0cm}%
\label{fig:testsetwhole}
\end{figure}

In order to test the feasibility of clustering methods,
we first complied several test sets of superconductors from literature 
\cite{poole-book,hirsch15}, each consisting of 25--40 materials. 
Such a small number of entries allows us to look into each superconductor
individually and verify its cluster assignment. In the first set, we included
a variety of different superconductors from literature,
including elements, cuprates, pnictides, heavy fermions, transition metal 
dichalcogenides, Chevrel compounds, etc. 
In Fig.~\ref{fig:testsetwhole} we display the results of 
clustering analysis and plot them using t-SNE. The two axes (Y$_1$ and Y$_2$)
are the reduced dimensions of the original data set; they do not have any
physical meaning. One can readily 
identify three major clusters, which correspond to cuprates (red circles),
pnictides (green circles) and what we previously \cite{roter20} 
called "others" (blue circles). The three
clusters are well separated from each other, and no data points
can be found between them. It is also possible to identify 
some sub-cluster structures in Fig.~\ref{fig:testsetwhole}.
However in our experience, it is better to do clustering in steps:
once we identify the three main clusters, we cluster each of 
them separately. 

It is interesting to notice that in Fig.~\ref{fig:testsetwhole} 
oxide superconductors Sr$_2$RuO$_4$, B$_{0.6}$K$_{0.4}$BiO$_3$, 
LiTi$_2$O$_4$ and the recently discovered \cite{li19} nickelate 
Nd$_{0.8}$Sr$_{0.2}$NiO$_2$ are clustered with the cuprates. This 
should not be surprising, as they are chemically much more similar to 
cuprates (copper-oxides) than to any other group of superconductors
shown in Fig.~\ref{fig:testsetwhole}.


\begin{figure}[tbp]
\vspace*{0.0cm}%
\centerline{\includegraphics[width=4in]{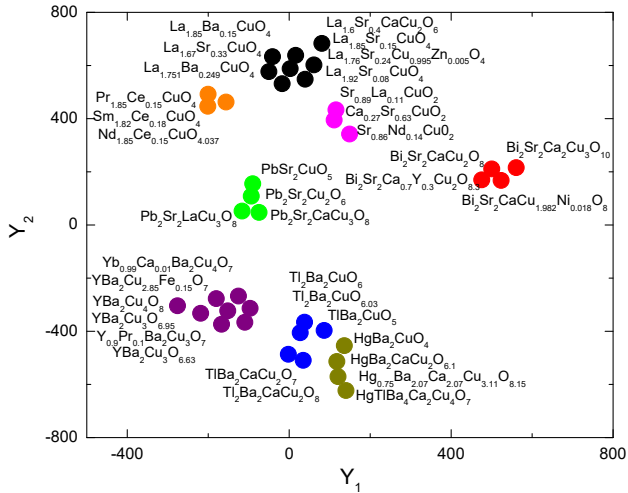}}%
\vspace*{-0.25cm}%
\caption{(Color online).  t-SNE plot of the 
cuprate test set consisting of 39 cuprate superconductors. 
Machine learning clustering identified eight major cuprate families:
BSCCO (red), LSCO (black), YBCO (purple), NCCO (orange), 
Hg-based (dark yellow), Tl-based (blue), Pb-based (green)
and infinite layer (magenta).}
\vspace*{0.0cm}%
\label{fig:testsetcuprate}
\end{figure}

The second test set includes only cuprates \cite{poole-book,hirsch15,chu15} 
and has 39 materials. Fig.~\ref{fig:testsetcuprate} shows the results of 
clustering, where one can identify eight major cuprate families: 
BSCCO (red), LSCO (black), YBCO (purple), NCCO (orange), 
Hg-based (dark yellow), Tl-based (blue), Pb-based (green)
and infinite layer (magenta). We notice that the hole-doped LSCO 
family is separated from the electron-doped NCCO family, even though
they are chemically and crystallographically very similar and are 
often assumed to belong to the same family \cite{poole-book}. 
Similarly, Tl-based
and Hg-based cuprates are considered to be in the same family, 
but the algorithm was able to separate. Not surprisingly, 
they are located close to each other in Fig.~\ref{fig:testsetcuprate}.


\begin{figure}[tbp]
\vspace*{0.0cm}%
\centerline{\includegraphics[width=4in]{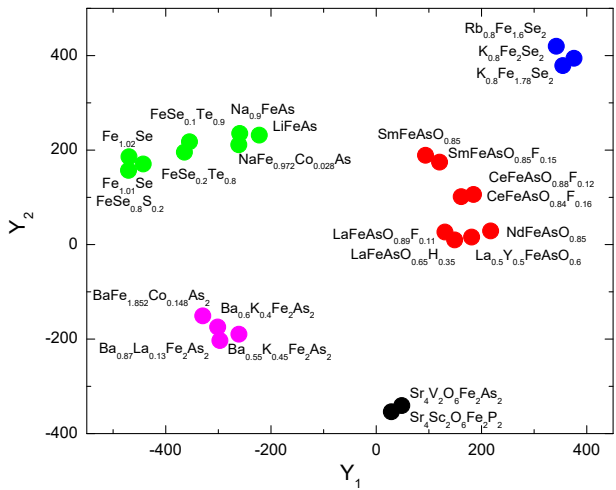}}%
\vspace*{-0.25cm}%
\caption{(Color online).  t-SNE plot of the 
pnictide test set consisting of 25 pnictide superconductors. 
Machine Learning clustering was able to identify five major families:
11/111 (green), 122 (magenta), 1111 (red), 245 (blue), 
and 42622 (black).}
\vspace*{0.0cm}%
\label{fig:testsetpnictide}
\end{figure}

The last test set contains 25 pnictide superconductors \cite{hosono15,ido21}. 
The results of clustering are shown in Fig.~\ref{fig:testsetpnictide}. 
One can identify five major pnictide families: 11/111 (green), 
122 (magenta), 1111 (red), 245 (blue), and 42622 (black). All five
families are well separated, with no data points in between clusters. 
We notice that there might be some sub-cluster structure within the 
11/111 family, which might be resolvable with a larger data set.


\section{Clustering SuperCon Database}

The results presented in the previous section (Figs.~\ref{fig:testsetwhole}, 
\ref{fig:testsetcuprate} and \ref{fig:testsetpnictide}) indicated
that clustering methods were capable of distinguishing between
different types of superconductors. 
We then proceeded to cluster the SuperCon database \cite{supercon}. After removing 
incorrect, incomplete and multiple entries, we were left with about 16,000 
superconducting materials. 
In order to improve the learning process, we removed entries with eight 
or more elements. We also note that the number of pnictide entries 
in the database is
much smaller compared with cuprates and others. Our calculations indicate
that clustering works better when the number of constituents in each 
cluster is similar. Therefore, for our calculations we randomly selected
between 1,400 and 1,500 superconductors of each type, for a total 
of approximately 4,500 entries.

\begin{figure}[tbp]
\vspace*{0.0cm}%
\centerline{\includegraphics[width=4in]{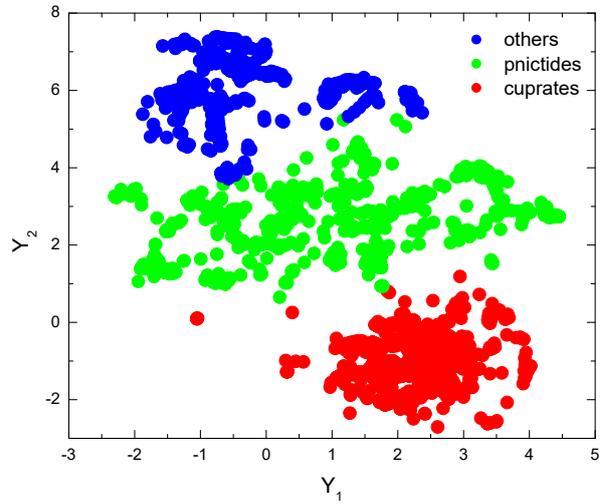}}%
\vspace*{-0.25cm}%
\caption{(Color online). t-SNE plot of 4,500 randomly selected
superconductors from the SuperCon database.
Three clusters are clearly seen, and they correspond to cuprate (red),
pnictide (green) and other (blue) superconductors.}
\vspace*{0.0cm}%
\label{fig:superconwhole}
\end{figure}

In Fig.~\ref{fig:superconwhole} we show the results of clustering
applied to this data set. Similar to Fig.~\ref{fig:testsetwhole},
one can identify three major 
clusters, which correspond to cuprates, pnictides and others. 
However, unlike clustering achieved for the test set in 
Fig.~\ref{fig:testsetwhole}, the separation of clusters is not 
as clear, and there are points between the clusters. 
This is not surprising considering the size and quality of SuperCon 
database.


\begin{figure}[tbp]
\vspace*{0.0cm}%
\centerline{\includegraphics[width=4in]{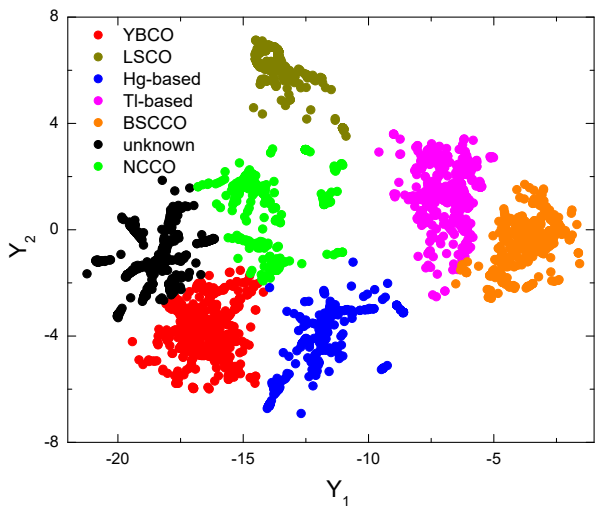}}%
\vspace*{-0.25cm}%
\caption{(Color online). t-SNE plot of cuprates.
Seven clusters are clearly identified, and they correspond
to the following families: YBCO (red), LSCO 
(dark yellow), BSCCO (orange), Hg-based (blue), Tl-based (magenta),
NCCO (green) and a cluster (black) that contains a variety of 
different cuprates, which the program was unable to cluster with
any other family.}
\vspace*{0.0cm}%
\label{fig:superconcuprate}
\end{figure}

After that, we extracted cuprates from the SuperCon database
as superconductors containing both copper and oxygen. 
The results of clustering are shown in Fig.~\ref{fig:superconcuprate}.
We can identify seven clusters, which correspond to YBCO (red), LSCO 
(dark yellow), BSCCO (orange), Hg-based (blue), Tl-based (magenta),
NCCO (green) and a cluster (black) that contains a variety of 
different cuprates, which the program was unable to cluster with
any other family. Manual backward validation indicates that this
cluster contains mostly cuprates with rare-earth elements 
like lutetium, dysprosium, erbium, holmium, etc. 
We also notice that, unlike our test set in Fig.~\ref{fig:testsetcuprate},
we are not able to identify the Pb-based and infinite layer families
in Fig.~\ref{fig:superconcuprate}, which we speculate 
is because of only a small number of their entries existing in the database.


\begin{figure}[tbp]
\vspace*{0.0cm}%
\centerline{\includegraphics[width=4in]{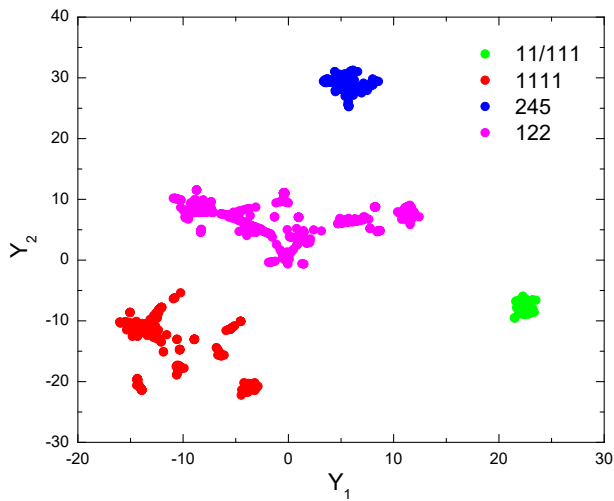}}%
\vspace*{-0.25cm}%
\caption{(Color online). t-SNE plot of 1,400 pnictides from
the SuperCon database.
Four clusters are clearly identified, and they correspond
to the following families: 11/111 (red), 122 (black), 1111 (green)
and 245 (blue).}
\vspace*{0.0cm}%
\label{fig:superconpnictide}
\end{figure}

Next, we extracted pnictides from the SuperCon database as 
entries that have iron and at least one of the following:
arsenic, phosphorous, selenium or tellurium. This resulted in 
a set with only about 1,400 entries. In Fig.~\ref{fig:superconpnictide} 
we show the result of clustering. 
The t-SNE plot indicates that there are four well separated
clusters: 11/111 (red), 122 (black), 1111 (green)
and 245 (blue). We notice that the 11/111 and 245 clusters
are much more compact compared to 122 and 1111 clusters. 
This might be due to the fact
that there are significantly fewer entries of 11/111 and 245 
superconductors in the SuperCon database. It is also clear that
there is some sub-cluster structure in the 122 and 1111 clusters,
which might be resolvable with a larger data set. 
ML clustering was not able to resolve the 42622 family, 
as there were very few entries of them in the database.


After removing cuprates and pnictes, we were left with approximately
7,000 superconductors, which we refer to as others. This is a very
diverse set of materials, and clustering it is not an easy task. 
As mentioned above, for most clustering algorithms the number of
desired clusters is an input parameter. However, for other 
superconductors we do not {\it a priori} know the number of clusters. 
As a result, we used DBSCAN algorithm which does not require the 
number of clusters to be pre-specified \cite{dbscan-comm}. 

\begin{figure}[tbp]
\vspace*{0.0cm}%
\centerline{\includegraphics[width=4in]{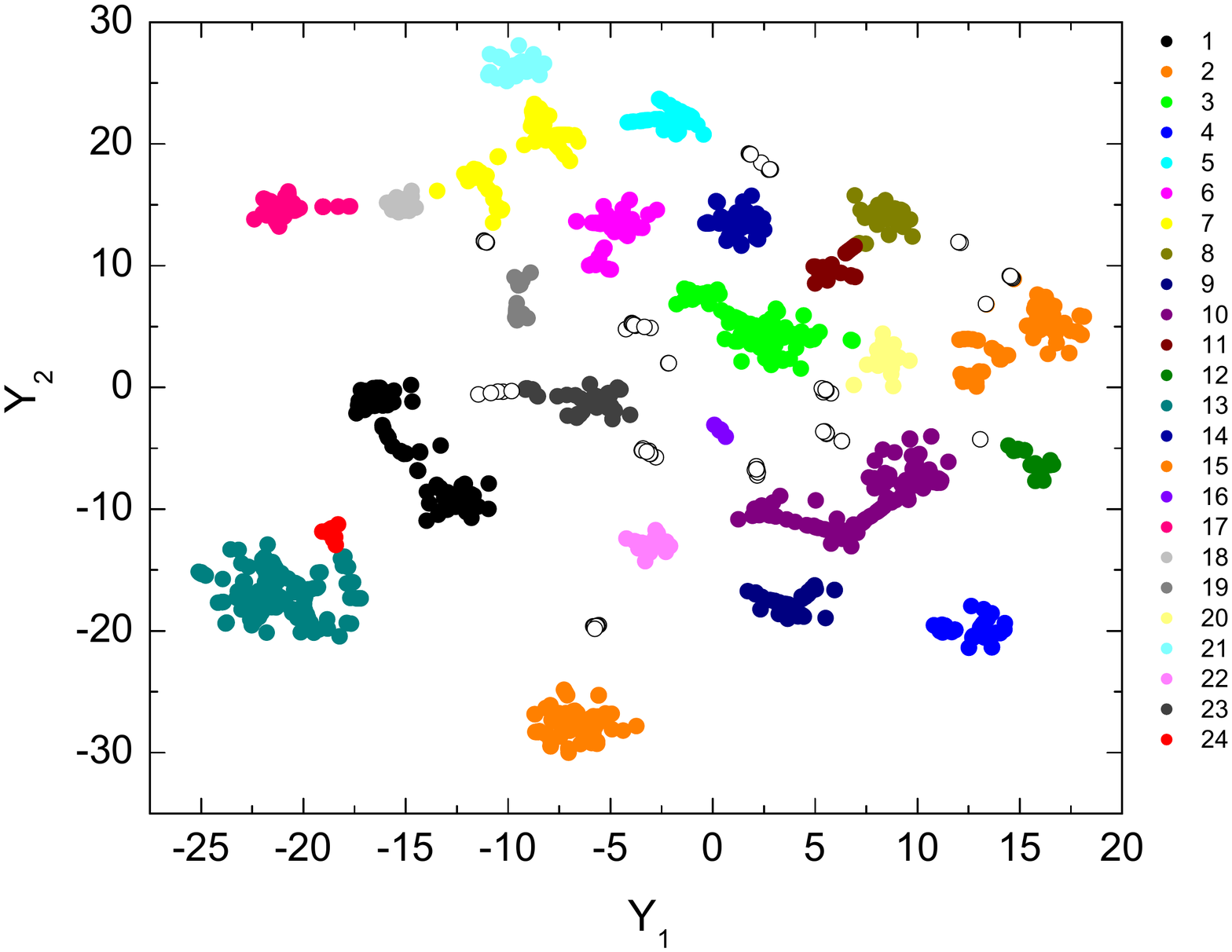}}%
\vspace*{-0.25cm}%
\caption{(Color online). t-SNE plot of other supercondutors from
the SuperCon database. DBSCAN was used for clustering, as it does
not require the number of clusters to be specified. The program 
separated the data set into 24 clusters. We were able to identify
them, and they are listed in Table.~\ref{table:others}. DBSCAN 
was also able to identify several outliers, which are shown with 
open circles. }
\vspace*{0.0cm}%
\label{fig:superconother}
\end{figure}

The results of calculations on other superconductors are shown in 
Fig.~\ref{fig:superconother}. 
The algorithm separated this data set into 24 clusters. 
We were able to identify them and they are listed
in Table~\ref{table:others}. For each cluster several representative
examples are shown. We notice that in each cluster there is 
predominant element or a group of elements, which is the 
common thread for that particular cluster. This is not surprising,
as the chemical composition was used as the only predictor in 
these calculations. 

DBSCAN also identify a number of outliers among other superconductors,
shown with open circles in Fig.~\ref{fig:superconother}. These 
points are usually separated from the main clusters. Some outliers are:
BaHg, LiHg$_3$, Ni$_{0.3}$Th$_{0.7}$, Th$_{0.9988}$U$_{0.0012}$, Cr$_{0.75}$Ru$_{0.25}$, 
Co$_{0.96}$Mo$_{0.04}$U$_6$, Co$_{0.94}$Rh$_{0.06}$U$_6$, Gd$_{0.001}$Th$_{0.999}$, 
Ag$_{0.7}$Zn$_{0.3}$, Ag$_{0.625}$Al$_{0.375}$, etc. 
In our opinion, these materials, and in particular their derivatives, 
should be explored further as seeds for potentially new classes of 
superconducting materials.


\begin{table*}   
\centering
\caption{Clusters identified in Fig.~\ref{fig:superconother} using DBSCAN. 
Several examples of superconductors from each cluster are shown in the last column.}
\begin{tabular}{ |c|l|l| } 
 \hline
 Cluster number & Common thread & Examples \\ 
 \hline
 1 & Zr- and Ti-based  & Zr$_2$Co$_{0.9}$Cu$_{0.1}$, Zr$_{0.8}$Mo$_{0.2}$, Ir$_{0.1}$Ti$_{0.9}$ , Ti$_{0.576}$V$_{0.384}$Ru$_{0.04}$ \\ 

 2 & V-based  & PtV$_{3}$, Ga$_{0.3}$V$_{0.7}$, V$_3$Si$_{0.6}$Sn$_{0.4}$ \\ 

 3 & Si-based  & SrGa$_{0.5}$Si$_{1.5}$, CaAl$_{0.8}$Ga$_{0.2}$Si, Lu$_{2}$Ir$_{3}$Si$_{5}$ \\ 
 
 4 & S-based  & Cu$_{1.5}$Mo$_{4.5}$S$_{6}$, NbPbS$_{3}$, Ni$_{0.02}$TaS$_{2}$ \\ 

 5 & In-based  & CeIn$_{3}$, ThIn$_{1.5}$Sn$_{1.5}$, Bi$_{0.343}$In$_{0.657}$ \\ 

 6 & Pd-based  & Li$_2$Pd$_{3}$B, As$_{0.5}$Ni$_{0.06}$Pd$_{0.44}$, ZrPd$_{2}$Al \\ 

 7 & Te-based  & IrTe$_{3}$, Pb$_{0.985}$Tl$_{0.015}$Te, Ir$_{0.95}$Pt$_{0.05}$Te$_2$ \\ 

 8 & La-based  & Bi$_3$La$_{4}$, La$_{3}$Rh$_{2}$Ge$_2$, AlGd$_{0.003}$La$_{2}$ \\ 

 9 & Se-based  & TiSe$_{2}$, Sr$_{0.2}$WSe$_{2}$, La$_{2.85}$Pr$_{0.15}$Se$_4$ \\ 

 10 & C-based  & C$_{1.35}$La, Cs$_{2}$RbC$_{60}$, La$_{0.2}$Th$_{0.8}$NiC$_2$ \\ 
 
 11 & Pt-based  & SrPt$_3$P, Pt$_{5}$Th, PrPt$_{2}$B$_{2}$C \\ 

 12 & Ni-based  & GdSmNi$_{2}$, HoNi$_{2}$B$_{2}$C, MgC$_{1.25}$Ni$_{3}$ \\ 

 13 & Nb-based  & Nb$_{3}$Sb, AlGeNb$_{3}$, Nb$_{4}$FeSi \\ 

 14 & Ge-based  & ThPt$_2$Ge$_{2}$, Pr$_{0.5}$Eu$_{0.5}$Pt$_4$Ge$_{12}$, CeNiGe$_{3}$ \\ 

 15 & B-based  & MgB$_{2}$, DyRh$_{2}$Ir$_{2}$B$_4$, Ca$_{3}$Rh$_{8}$B$_2$ \\ 

 16 & Be-based  & LuBe$_{13}$, Be$_{0.9}$Ni$_{0.1}$, MgReBe$_{12}$ \\ 

 17 & Oxides    & Ba$_{0.19}$K$_{0.81}$BiO$_{3}$, Ag$_5$Pb$_{2}$O$_{6}$, Bi$_{4}$S$_{2.91}$Se$_{0.09}$O$_4$ \\ 

 18 & Bi-based  & Bi$_{0.6}$Tl$_{0.4}$, Ba$_{2}$Bi$_{3}$, Bi$_{0.525}$Pb$_{0.32}$Sn$_{0.155}$ \\ 

 19 & Au-based  & Au$_{0.84}$In$_{0.16}$, AlScAu$_{2}$, Au$_{0.85}$Pd$_{0.15}$Ga$_2$ \\ 

 20 & Ir-based  & CaIr$_{2}$, SrIr$_{2}$As$_{2}$, Ir$_{0.8}$Pt$_{0.2}$ \\ 

 21 & Sn-based  & NbSn$_{3}$, Ca$_{2.25}$Sr$_{0.75}$Rh$_4$Sn$_{13}$, RhSn$_{2}$ \\ 

 22 & Ta-based  & Hf$_{0.4}$Ta$_{0.6}$, Be$_{2}$Ta$_{3}$, Nd$_{1.72}$Ta$_{3.28}$S$_2$ \\ 

 23 & Ru-based  & Ce$_{0.85}$La$_{0.15}$Ru$_{2}$, Ir$_{0.43}$Ru$_{0.57}$, Cr$_{0.5}$Ru$_{0.5}$ \\ 

 24 & NbN-based  & NbC$_{0.5}$N$_{0.47}$, Nb$_{2}$BN, NbN$_{0.87}$O$_{0.13}$ \\ 
 \hline
\end{tabular}
\label{table:others}
\end{table*}



\section{Summary}

We have made the first attempt at clustering superconductors
using ML methods. It was shown that ML clustering can achieve, 
and in some cases (cuprates, for example) exceeded human level 
performance. Clustering 
was done based on chemical composition only. We speculate that
when additional physical parameters are included as predictors, even better results 
might be obtained. In particular, one should consider empirical predictors that are 
believed \cite{matthias} to have strong relations with superconductivity,
such as crystal symmetry, layered structure, density of states at the Fermi level,
the number of valence electrons,
etc. We showed that the best results are achieved when superconducting
materials are clustered in steps. A similar approach might be useful in 
other data-driven materials calculations.  

We also discussed the limiting factors, i.e. wrong entries in the database. 
One can hope that with a more reliable database, which will also include 
more predictors, better clustering results can be achieved. In that regard, 
the recently initiated SuperMat project \cite{supermat}, which uses
text mining and other natural language processing techniques to construct
a superconductor database, might turn out to be fruitful. 



%
%

\begin{thebibliography}{00}
\bibitem{dbscan-comm} DBSCAN requires two input parameters to be specified: $\epsilon$ and minPts. 
As has been pointed out before, specifying these two parameters is not trivial \cite{wunsch-book}. 
Their values determine both the number and content of clusters. In recent years, there have 
been numerous attempts to algorithmically determine the values of $\epsilon$ and minPts 
\cite{schubert17,rahmah16,starczewski20}. However, as stated above, in our experience the 
best results when clustering 
superconductors are obtained when machine learning techniques are used in concert with human knowledge 
of superconductors. 
Therefore, we have manually adjusted the values of $\epsilon$ and minPts until the most 
physically meaningful clusters were obtained. The values used for clustering in 
Fig.\ref{fig:superconother} were $\epsilon$ = 3 and minPts = 30 and the distance function 
was cityblock.
\end{thebibliography}

\end{document}